\documentclass[a4paper,10pt]{article}

\usepackage{amssymb}
\usepackage{amsmath}
\usepackage[usenames,dvipsnames]{color}
\usepackage{hyperref}
\usepackage[left=.8in,right=.8in,top=.8in,bottom=.8in]{geometry}                  

\hypersetup{
    colorlinks=true,         
    linkcolor=blue,          
    citecolor=red,        
    urlcolor=Violet             
}

\newtheorem{theo}{Theorem}

\newcommand{\mean}[1]{\ensuremath{\lf\langle #1 \rt\rangle }}
\newcommand{\diby}[2]{\ensuremath{\frac{\delta #1}{\delta #2}}}

\def\be{\begin{equation}}
\def\ee{\end{equation}}
\def\bea{\begin{eqnarray}}
\def\eea{\end{eqnarray}}

\def\lf {\ensuremath{\left}}
\def\rt {\ensuremath{\right}}

\title{A construction principle for ADM-type theories in maximal slicing gauge}
\author{\bf Henrique Gomes\footnote{\href{mailto:gomes.ha@gmail.com}{gomes.ha@gmail.com}}\\\it  Department of Physics,  University of California, Davis,   CA, 95616}

\begin{document}

\maketitle

\begin{abstract}
  
The differing concepts of time in general relativity and quantum mechanics are widely accused as the main culprits in our persistent failure in finding a complete theory of quantum gravity. Here we address this issue by constructing ADM-type theories \emph{in a particular time gauge} directly from first principles. The principles are expressed as conditions on phase space constraints: we search for two sets of spatially covariant constraints, which generate symmetries (are first class) and gauge-fix each other leaving two propagating degrees of freedom. 
 One of the sets is the Weyl generator tr$(\pi)$, and the other is a one-parameter family containing the ADM scalar constraint $\lambda R- \beta(\pi^{ab}\pi_{ab}+(\mbox{tr}(\pi))^2/2))$.  The two sets of constraints can be seen as defining ADM-type theories with a maximal slicing gauge-fixing. This work provides an independent, first principles derivation of ADM gravity.  
  The principles above are motivated by a heuristic argument relying in the relation between symmetry doubling and exact renormalization arguments for quantum gravity, aside from compatibility with the spatial diffeomorphisms. As a by-product, these results address one of the most popular criticisms of Shape Dynamics: its construction starts off from the ADM Hamiltonian formulation.
 The present work severs this dependence: the set of constraints yield reduced phase space theories that can be naturally represented by either Shape Dynamics or ADM. More precisely, the resulting theories can be naturally ``unfixed" to encompass either spatial Weyl invariance (the symmetry of Shape Dynamics) or refoliation symmetry (ADM).  
 \end{abstract}

\section{Introduction}

It is difficult today for physicists to imagine space \emph{and} time as anything other than the amalgamated 
``space-time" of general relativity. 
  However, space-time itself remains largely unassailable to quantization. One of the reasons for this resistance is the incongruence between the  notions of time in quantum mechanics and general relativity, a problem that resurfaces in different guises for different approaches to quantum gravity.\footnote{For an interesting take on how this problem emerges in AdS/CFT for example, see  the lectures \cite{Nima-Lectures}.  } 

It is of interest then to have a different formulation of a gravity theory with the same observables as General Relativity, but which does not use so explicitly the merging of time and space. In this respect, a useful analogy could be made with quantum mechanics.
By the end of the XIXth century, the indeterministic character of quantum mechanics was still nowhere apparent in pre--existing physics. However, had the physicists of the time  somehow known beforehand that the notion of classical determinism was problematic, there would have been a straightforward way of investigating how its breaking might be reconciled with Newtonian mechanics. Namely, by using the least action principle's re--formulation of mechanics, one could have seen how not a single deterministic path was being taken by a given particle, but that many paths might be somehow involved  in  an observed `classical' reality. This slight disentanglement between the  observed classical behavior and the idea of  a unique path would have been a powerful clue in the direction of quantum mechanics.\footnote{The relation with Fermat's principle for the path of light rays and the wave-like behavior of light, would have been other available clues at the time.} 

In the same way, a Hamiltonian version of general relativity goes a long way into separating the notions of time and space, especially if seen on its own grounds and not in relation to the space-time picture.\footnote{For instance, not demanding positivity of the Lagrange multiplier $N$, usually assumed to be positive so that it can be related to the time-time component of the 4-metric  \cite{SD:Birkhoff}.} But as with the principle of least action, this single clue is not enough to point us into the right direction, since also in the Hamiltonian formulation refoliation invariance  figures prominently. One could then consider a gauge-fixed version of Hamiltonian ADM as the ``true theory", reformulating all the dynamics to be valid only in that single gauge. But how to choose one ``special gauge" over another? In this paper we will derive a preferred set of gauge-fixed theories from quite general symmetry principles. The ``special gauge" mentioned above, turns out to represent  spatial Weyl invariance.

These results complete a genuine first-principles disentanglement between the concepts of space-time and gravity, all the while -- much like the Least Action Principle wrt to classical mechanics -- retaining (most of) the physical observables of general relativity. We provide an independent construction of ADM gravity from first principles.  In the process, we find an independent construction principle for the reduced phase space version of Shape Dynamics. More precisely, we are obligatorily led to  gauge-fixed versions of ADM type theories that have a Weyl-invariant dual. The present work is thus an independent construction principle for the phase-space reduced theory representing both Shape Dynamics and ADM gravity. 

\subsubsection*{An independent construction principle for Shape Dynamics}

As will be made explicit in section \ref{sec:SD}, the construction of Shape Dynamics relies heavily on the constraint structure of Hamiltonian ADM, and also on the York procedure to solve the initial value problem of GR. This raises serious questions about the theoretical basis of Shape Dynamics: it might be an interesting description of phenomena that we perceive as gravity, but how could someone have discovered it if Einstein had not first discovered General relativity? In other words, is there a construction principle that one might invoke for Shape Dynamics that is independent of general relativity?

In the case of general relativity, there  is an abundance of construction principles that lead more or less uniquely to GR. The first of these, introduced by Lovelock \cite{Lovelock}, shows that the Einstein tensor is the unique generally covariant divergence-free tensor with 2 derivatives of the metric in 4-dimensions.  Lovelock's paper was published in 1970, already a good half of a century after the advent of general relativity.   From the Hamiltonian setting,  Hojman, Kuchar and Teitelboim (HKT) \cite{HKT} showed that a similar uniqueness emerged if one asked for phase space representations of the commutation algebra of  vector fields in a hypersurface orthogonal decomposition. Together with certain requirement on the number of derivatives,  the absence of unwanted terms (like our own absence of derivative couplings) and two propagating degrees of freedom, it was shown that the ADM constraint structure emerged rather uniquely. 

  From the first introduction of Shape Dynamics in \cite{SD_first}, ways to construct the theory in the same manner as done in \cite{HKT} were sought after, but ultimately failed. This failure was due mainly to the fact that the commutation algebra of spatial conformal diffeomorphisms is intrinsic to the hypersurface, and thus if one followed the HKT construction one  ``missed"   the emergence of the global Hamiltonian of Shape Dynamics (or of any time-evolution generator).
  
\subsubsection*{Our results}
  
   Here we will provide a mechanism that allows for the simultaneous emergence of a one-parameter family of Hamiltonian constraints (that include ADM),\footnote{Only in vacuum is it a one-parameter family. With the inclusion of matter this likely becomes at least a 2-parameter family. } and the spatial Weyl generator of Shape Dynamics. The principles imposed to obtain this result are very simple. We want two distinct sets of constraints, each compatible with the spatial diffeomorphism constraint and each generating local symmetries in phase space. But the crucial additional requirement is that one member of the pair gauge-fixes the other, leaving two  remaining  propagating degrees of freedom. Being slightly cavalier about the fine-print, we can state the requirements in one condensed soundbite: we want spatially covariant, mutually gauge-fixing symmetries that leave two propagating physical degrees of freedom.\footnote{Covariant in the sense that they are compatible with the action of the spatial diffeomorphisms.} 
  
  It is our view that the above first principles are simple and powerful enough to require no further justifications. Nonetheless, one can indeed find a further heuristic argument to justify this particular criterium in the search for a gravity theory: symmetry doubling \cite{SD:Sym_doub}. Symmetry doubling is the phenomena by which certain BRST gauge-fixed dynamical systems possess two complementing BRST symmetries (this is \emph{not} anti-BRST symmetry).  In \cite{SD:Sym_doub} it is shown that one of the  necessary conditions for symmetry doubling to occur is that the phase space possesses two sets of mutually gauge-fixing sets of first class constraints. This condition, together with the requirement that the system be ``pure constraints" - i.e. its initial Hamiltonian be a sum of constraints -   guarantees that the extended BRST gauge-fixed Hamiltonian possesses two complementing sets of BRST invariances, corresponding to the two symmetries generated by the first class constraints. 
  
  It is no secret that restricting theory space to obey certain symmetries is very advantageous for  renormalization, and it is particularly useful in the study of exact renormalization group flow in gravity \cite{Reuter}. The arguments for why symmetry doubling in particular is useful in this context is laid out more fully in \cite{Tim_effective}  and \cite{Astrid}. We take this as a starting point and look for the twin sets of constraints satisfying the conditions necessary for their BRST extensions to present symmetry doubling, being led almost inexorably to a unique set.
  
  \section{Preliminaries}

\subsection{Shape Dynamics}\label{sec:SD}

The first step in the construction of Shape Dynamics is to  write out the constraints of canonical GR in its 3+1 ADM form: 
\begin{eqnarray}
\label{equ:scalar constraint}S(x):= \frac{G_{abcd}\pi^{ab}\pi^{cd}}{\sqrt g}(x)-R(x)\sqrt g(x)=0\\
\label{equ:momentum constraint} H_a:={\pi^{a}_b}_{;a}=0
\end{eqnarray}
where the points $x$ belong to an open 3-manifold $\Sigma$, $g_{ab}$ is the spatial 3-metric and its conjugate momenta $\pi^{ab}$ (intimately related to the extrinsic curvature of a foliation). The scalar constraint \eqref{equ:scalar constraint} generates on-shell refoliations of spacetime, while the momentum constraint generates foliation preserving diffeomorphisms. The second step is to perform a canonical transformation in an extended phase space with coordinates $(g_{ab},\pi^{ab},\phi, \pi_\phi)$. The canonical transformation is of the form: 
$$(g_{ab},\pi^{ab},\phi,\pi_\phi)\mapsto (e^{4\psi}g_{ab},e^{-4\psi}\pi^{ab},\phi,\pi_\phi-4\pi)$$
where $\pi=g_{ab}\pi^{ab}$. The Stuckelberg -- extended action is invariant under this transformation. We have an extra first class constraint in this extended theory, which is generated by $\pi_\phi-4\pi\approx 0$. 

Under these transformations the scalar constraint \eqref{equ:scalar constraint} becomes, for $\phi=\ln\Omega$  
\be\label{LY} \nabla^2\Omega+R\Omega-\frac{1}{8}\pi^{ab}\pi_{ab}\Omega^{-7}=0 
\ee
Ignoring boundary terms (see \cite{Asymptotic_SD} for details on how to treat the boundary terms), the smeared diffeomorphism constraint becomes
\be\label{equ:conf_diffeo_constraint}
H_a(\xi^a)=\int_\Sigma \left(\pi^{ab}\mathcal{L}_\xi g_{ab}+\pi_\phi\mathcal{L}_\xi \phi \right) d^3 x
\ee

Now one performs the gauge-fixing $\pi_\phi=0$ on this extended system. The only constraint that is second class with respect to this gauge-fixing is exactly \eqref{LY}. This constraint can be solved for $\Omega$ \cite{York} and the system reduced to a system with the canonical Poisson brackets of the variables $(g_{ab},\pi^{ab})$. It turns out that one global Hamiltonian remains first class and thus is not gauge-fixed. There are different forms of expressing this Hamiltonian, one of them is as a total volume constraint \cite{SD:FAQ}:
\begin{equation}\label{equ:volumeConstraint}
 \int_\Sigma\sqrt{|g|}\left(1-e^{6\phi_o[g,\pi]}\right)\approx 0.
\end{equation}
This is the generator of the evolution in Shape Dynamics.

The other remaining first class constraints after phase space reduction are the spatial diffeomorphism constraint
$$\int d^3 x\left(\pi^{ab}\mathcal{L}_\xi g_{ab}\right)=0
$$
 and the  Weyl (or conformal) constraint $$\pi=0.$$

\subsection{Conformal 3+1 ADM in maximal slicing}

Shape Dynamics is intimately related to a gauge-fixing of ADM, namely, either a constant mean curvature (CMC) gauge fixing for the closed spatial manifold case\footnote{Here when we mention a constant-mean curvature gauge-fixing, we mean first of all not that the trace of the extrinsic curvature is equal to a single constant for all times, but that at each time it is a constant. Secondly, when we mention the gauge-fixing of ADM, we don't mean that we are given a space-time and then study its properties in a certain gauge. Instead, we impose the gauge on theory space itself, i.e. the dynamical equations of motion should all be restricted to this gauge. }, or the maximal slice gauge fixing for the open manifold case. The special property of these gauge fixings, respectively $\pi-\mean{\pi}\sqrt{g}=0$ and $\pi=0$, is that they also moonlight as generators of spatial Weyl transformations (in the CMC case it is the generator of total volume preserving Weyl transformations).

So why not work directly with the gauge-fixed version of ADM? The issue, as with any gauge-fixing, is that as with any gauge-fixed theory, the reduced variables are in general non-geometric, sometimes non-local, and in any case not easy to work with. Furthermore, to be able to reduce the theory, one must have auxiliary structure, present as a non-dynamical background metric with regards to which we parametrize the conformal factor of each metric. For the complete construction of the reduced theory see the appendix of \cite{SD:FAQ}. 

Nonetheless, the reduced theory has two gauge ``unfixings" which return us to the local geometric variables $(g_{ab},\pi^{ab})$: one is ADM itself, revealing refoliation symmetry. The other is Shape Dynamics, revealing Weyl symmetry. By finding this particular set of preferred gauge-fixed theories, we are thus finding the theory that can be naturally extended into either ADM or Shape Dynamics. 

\subsection{Symmetry doubling}\label{sec:sym_doub}
 The motivation for the argument we will use in the main part of the paper uses the core results of ``symmetry doubling". Before we briefly explain the concept, we should stress that the central result of this paper is independent of symmetry doubling, which we regard solely as the motivation heuristics behind the set of restrictions on the constraint space we consider.

In the case of gravity, symmetry doubling was applied to obtain doubly general relativity \cite{SD:Sym_doub}. Doubly general relativity involves a BRST treatment of a CMC gauge fixing of ADM. The general statement of symmetry doubling is that in the case of pure constraints theories - such as ADM - gauge fixing terms that are also symmetry generators possess a special role in the classical BRST formulation of the gauge-fixed theory: the gauge-fixed Hamiltonian in this instance has the BRST symmetries related to both the original symmetry and  that of the gauge-fixing term.
 In the remaining of this section we largely copy sections 4.2 and 4.3 of \cite{SD:Sym_doub}.

 For a rank one BRST charge, related to the constraints $\chi_a$ with structure functions $U_{ab}^c$ we have
\be   \Omega= \eta^a\chi _a-\frac{1}{2}\eta^b\eta^aU_{ab}^c P_c
\ee
where $\eta^a$ are the ghosts associated to the constraint transformations, and $P_b$ the canonically conjugate ghost momenta. The \emph{rank} of a system can be identified with the order of ghost momenta required for constructing a nilpotent BRST charge.

The gauge-fixed Hamiltonian is constructed by choosing a ghost number $-1$ fermion $\tilde\Psi=\tilde\sigma^\alpha P_\alpha+...$, where $\{\tilde\sigma^\alpha\}_{\alpha \in \mathcal A}$ is a set of proper gauge fixing conditions. 
Denoting the BRST invariant extension of the on-shell Hamiltonian (where all constraints are set to vanish) by $H_o$, the general gauge fixed BRST-Hamiltonian is written as
\begin{equation}
 H_{\tilde\Psi}=H_o+\eta^\alpha V_\alpha^\beta P_\beta+\{\Omega,\tilde\Psi\},
\end{equation}
where $\{H_o,\chi_\alpha\}=V_\alpha^\beta\chi_\beta$ and the bracket is extended to include the conjugate ghost variables. The gauge fixing  changes the dynamics of ghosts and other non-BRST invariant functions, but maintains evolution of all BRST-invariant functions. The crux of the BRST-formalism is that the gauge-fixed Hamiltonian $H_{\tilde\Psi}$ commutes strongly with the BRST generator $\Omega$. Although  gauge symmetry is completely encoded in the BRST transformation $s  := \{\Omega, . \}$, and we have fixed the gauge, the system retains a notion of gauge-invariance through BRST symmetry.

Applying this to a generally covariant theory, i.e. a system with vanishing on-shell Hamiltonian $H_o=0$, we find that the gauge-fixed BRST-Hamiltonian takes the form
\begin{equation}
 H_{\tilde\Psi}=\{\Omega,\tilde\Psi\}.
\end{equation}
Now comes the rather simple central insight that makes symmetry doubling possible: if the set $\{\sigma^\alpha\}_{\alpha \in \mathcal A}$ is both a proper gauge fixing for $\chi_\alpha$ and a first class set of constraints, one can construct a nilpotent gauge-fixing $\Psi$ analogous to the construction of the BRST generator; the only difference is that ghosts and antighosts are swapped.  The result for rank one theories is given simply by
\be \Psi=\sigma^\alpha P_\alpha-\frac{1}{2} P_b P_aC^{ab}_c \eta^c
\ee

This means that the Hamiltonian is invariant under two BRST transformations
\begin{equation}
 \begin{array}{rcl}
   s_1 . &=& \{ \Omega , . \} \\
   s_2 . &=& \{ . , \Psi \} ,
 \end{array}
\end{equation}
which follows  directly from the super-Jacobi identity and nilpotency of both $\Omega$ and $\Psi$. It should be emphasized that one of the strengths of the BRST treatment is that such symmetries are not restricted to be on-shell, but take effect all over phase space. The dual structure of the gauge-fixing fermion with respect to the initial system, and the double invariance of the gauge-fixed Hamiltonian, allows us to see rather straightforwardly that  the two symmetries act at the same level.\footnote{There is a more complicated question of how these invariances act on the observables. For this we point to \cite{Tim_effective}. }

\section{An independent construction principle}\label{sec:Independence}

\subsection{Setting}

In this section we present an argument for why, in (nearly) the words of Einstein, `God might have had very little choice in constructing Shape Dynamics'. In fact, what we will argue for is that renormalization group flow arguments might favor the gauge-fixed theory of ADM in CMC. As we explained in the introduction, ADM in CMC has a non-local flavor to it  (at least non-geometrical), and it contains reference to a fiducial fixed metric. To obtain a local description we must introduce gauge-redundancy (which is always the way gauge-freedom is introduced in any case). This introduction 
can be made in two ways: either through Weyl symmetry or through refoliation symmetry.


Our method here will be to  look for two  constraints, i.e. functionals in the phase space of gravity $(g_{ab},\pi^{ab})$  that i) are scalar ii) generically gauge-fix each other, and iii) are first class with respect to the diffeomorphism constraint. The first class property ensures that they generate symmetries themselves, the gauge-fixing property with respect to each other (generically second class) ensures that they will form a symmetry doubling pair, and finally, being scalar constraints ensures that the resulting theory has two propagating degrees of freedom. The ``generic" property mentioned above guarantees that there will be no obstructions to the flow in the hypothetical critical surface. 

\paragraph*{The restrictions in the space of possible constraints}

First and foremost we emphasize the way in which we restrict our  search. We demand of our candidate terms the following: 
\begin{itemize}
\item The constraints must be scalar. That is, they must represent one degree of freedom per space point (so that we obtain a physical theory of two degrees of freedom per space-point). 
\item Individually, each set  must be first class when taken in conjunction with the spatial diffeomorphism constraint (which we take to be a fundamental symmetry of our description). The first class requirement is present so that the constraints can be taken to generate symmetries, and have an associated BRST charge. 
\item We will look for two sets of constraints that  are generically second class  with respect to each other. This just means that each will serve as a good gauge-fixing for the other.  
\end{itemize}
To sum up, we are looking for two symmetries that gauge-fix each other. It is remarkable that such minimal assumptions generate the strong results we present below. Indeed, we need further assumptions in our theory space. The following are our working assumptions, although our position is that they might be gotten rid of or replaced by more natural conditions in the future. 
\begin{itemize}
\item The terms should depend on both the metric and the momenta (so that they include time and are not  purely intrinsic to the hypersurface geometry).
\item There should be no derivative coupling terms (in phase space these are represented by terms like $R_{ab}\pi^{ab}$, i.e. terms that mix two spatial derivatives and time derivatives). 
\item We will look at all terms that lie in this category, up to fourth derivatives. For higher than fourth derivatives we only include the exponents of the scalar curvature $R^n$. 
\end{itemize}

\subsection{Commutation (first-class) properties of the constraints}\label{sec:first_class}

 We start by defining the following set of functionals obeying the restrictions above: 
\be\label{def:A} A(\alpha,\beta,\gamma,a,b,c):= \alpha R^{ab}R_{ab}+\beta \nabla_a\nabla_bR^{ab}+ \gamma\nabla^2R+\mu_n R^n +\frac{a\pi^{ab}\pi_{ab}+b\pi^2}{g}+ c\frac{\pi}{\sqrt g}
\ee
where  $N_1$ and $N_2$ are smearing functions. Our aim in this section is to first restrict the possible functionals above by the demand that 
 the following Poisson bracket be weakly zero:
 \be\label{PB_A}\{A(N_1),A(N_2)\}\approx 0
\ee
weak equality means that the right hand side of \eqref{PB_A} vanishes whenever $\pi^{ab}_{~;b}=A=0$. 
We are not being as general as we could in the definition of $A$ in \eqref{def:A}. As will become clear from the calculation, we could add terms  with arbitrary powers of momenta, such as $$\frac{d\pi^a_c\pi^c_b\pi^b_a+e\pi^{ab}\pi_{ab}\pi+f\pi^3}{g^{3/2}}$$ and so on, without altering the result. Our main result is Theorem \ref{theo:prop}.

For the calculation, we first  need some preparatory results. For the  variations we will leave out terms that do not include derivatives of the Dirac delta tensors, since by commutativity these terms will not contribute to the Poisson brackets in \eqref{PB_A}. We will denote this equality up to linear terms by a dot over the equal sign. The variations are:
\begin{eqnarray}
\label{equ:Ricci_var}
\diby{R_{cd}(x)}{g_{ab}(y)}&=&-\frac{1}{2}\left(\delta^{(ab)}_{(cd)}\nabla^2\delta(x,y)+g^{ab}\delta(x,y)_{;cd}
-g^{ef}\left(\delta^{(ab)}_{(fc)}\delta(x,y)_{;de}+\delta^{(ab)}_{(fd)}\delta(x,y)_{;ce}\right)\right)\\
\label{equ:R_var}
\mu_1:~~\diby{R(x)}{g_{ab}(y)}&\dot{=}&-g^{ab}\nabla^2\delta(x,y)+\delta(x,y)^{;ab}\\
\beta:~~\diby{(\nabla_c\nabla_d R^{cd}(x))}{g_{ab}(y)}&=&-\frac{1}{2}\Big[(\nabla^2\delta(x,y))^{;ab}+g^{ab}(\delta(x,y)_{;cd})^{cd}-(\delta(x,y)_{;c})^{abc}-(\delta(x,y)_{;c})^{acb}\nonumber\\
&~& +2R^{c(a}{\delta(x,y)^{;b)}}_{;c}-R^{cd}\delta(x,y)_{;cd}g^{ab}+R^{ab}\nabla^2\delta(x,y)+4{R^{c(a}}_{;c}\delta(x,y)^{;b)}\nonumber\\
&~&-2{R^{cd}}_{;d}\delta(x,y)_{;c}g^{ab}+2R^{c(a;b)}\delta(x,y)_{;c}-R^{ab;c}\delta(x,y)_{;c}\Big]\label{equ:nabla_Ricc_var}\\
\gamma:~~\diby{\nabla^2R(x)}{g_{ab}(y)}&\dot{=}& -\nabla^2(\nabla^2\delta(x,y))g^{ab}+\nabla^2(\delta(x,y)^{;ba})-R^{ab}\nabla^2\delta(x,y)\nonumber\\&~&-2R^{ab;c}\delta(x,y)_{;c}-R^{;a}\delta(x,y)^{;b}+\frac{1}{2}R^{;c}\delta(x,y)_{;c}g^{ab}
\end{eqnarray}
Lastly, upon contraction with $R^{cd}$ with \eqref{equ:Ricci_var} it is easy to see that:
\be\label{equ:Ricci_var_contracted}
\alpha:~~R^{cd}\diby{R_{cd}(x)}{g_{ab}(y)}=-\frac{1}{2}\left(R^{ab}\nabla^2\delta(x,y)+g^{ab}R^{cd}\delta(x,y)_{;cd}
-2R^{c(b}{(\delta(x,y)^{;a)})_{;c}}\right)
\ee

It is now easy to see that the Poisson bracket between any of the metric terms $(\alpha,\beta, \gamma)$ with $(a,b,c)$ is proportional to the contraction of \eqref{equ:R_var}, \eqref{equ:Ricci_var_contracted} and \eqref{equ:nabla_Ricc_var} with (the appropriately densitized)  $2a\pi_{ab}, 2b\pi g_{ab}, c g_{ab}$ respectively.\footnote{With the $(d,e,f)$ terms contraction would be with $(3d\pi_{bc}\pi_{a}^c, e(2\pi_{ab}\pi+\pi^{cd}\pi_{cd}g_{ab}), 3f\pi^2 g_{ab})$ respectively.}

Let us start, for illustration, with the $(\mu_1, a,b,c)$ term:
\begin{eqnarray}
\{A(\mu_1,a,b,c)(N_1),A(\mu_1,a,b,c)(N_2)\}&=&\mu_1\int d^3 x  N_2\left(-\nabla^2N_1g^{ab}+N_1^{;ab}\right)\left(2a\pi_{ab}+ 2b\pi g_{ab}+ c g_{ab}\right) -(N_1\leftrightarrow N_2)\nonumber\\
&=&\mu_1\int d^3 x N_2\left(-\nabla^2N_1((2a+4b)\pi+2c)  +2a N_1^{;ab}\pi_{ab}\right)-(N_1\leftrightarrow N_2)\label{equ:nuabc}
\end{eqnarray}
This term already presents many of the features we will use in the other calculations. First, note that the last term $ \int N_2(N_1^{;ab}\pi_{ab})-(N_1\leftrightarrow N_2)$ can be set proportional to the diffeomorphism constraint and thus vanishes on-shell. The other terms cannot be set proportional to $A(\mu_1,a,b,c)$  nor to some of the higher order terms $\alpha$, $\beta$, $\gamma$, $\mu_n$, nor to the diffeomorphism constraint, unless $a=-2b , c=0$ (no imposition on $\mu_1$ so far). \footnote{Note for instance, that the term $-\nabla^2N_1((2a+4b)\pi$ is of course proportional to $\pi$ (i.e. the $c$ term), but there are no respective terms that come with the Ricci scalar (the $\mu_1$ term).} Furthermore, they cannot be canceled by the commutation of higher order terms, which possess higher derivative and no terms which contain no curvature, as does $-\nabla^2N_1((2a+4b)\pi+2c)$. 
So in order for the constraints to commute for $\mu_1\neq 0$,  we already know at least that $a=-2b$ and $c=0$. Thus we see the main reason to start with the $\mu_1\neq 0$ term, is that due to there being only one scalar formed from 2nd derivatives of the metric, we can draw conclusions regarding the coefficients  without reference to the higher order terms.

The commutation of the term $A(\mu_2,a,b,c)$ is exactly the same with the substitution $N_1\rightarrow 2R N_1$ in the first term: 
\begin{equation}\label{equ:muabc}
\{A(\mu_2,a,b,c)(N_1),A(\mu_2,a,b,c)(N_2)\}=2\mu_2\int d^3 x N_2\left(-\nabla^2(RN_1)((2a+4b)\pi+2c)  +2a (RN_1)^{;ab}\pi_{ab}\right)-(N_1\leftrightarrow N_2)
\end{equation} 
Now the term $-\nabla^2(RN_1)$ could be expanded, and we would obtain terms both in the $\mu_1$ set and in the $\gamma$ set. However non-zero terms lying in the  $\mu_1$ set of course demand  that $A$ have $\mu_1\neq 0$. However, as we have seen, the lower order terms of the $\mu_1$ set  cannot be canceled irrespective of the higher coefficients, unless, as we saw before, $a=-2b$ and $c=0$. However, unlike the case of $\mu_1$, this does not set \eqref{equ:nuabc} to a term proportional to the diffeomorphism constraint.  The terms $N_2(RN_1)^{;ab}\pi_{ab}-(N_1\leftrightarrow N_2)$ are no longer weakly zero, and thus disallows $\mu_2$. The same analysis follows for all powers of $R$.

Now we move on to the more complicated $\alpha, \beta, \gamma$ terms. The strategy to tackle these terms will be to first see what happens with the $a$ term. We will see that, irrespective of what happens with anything else, we cannot get rid of contractions of the Ricci curvature with the momenta (even after enforcing the momentum constraint), which cannot appear and thus sets either $a=0$ or $\alpha=\beta=\gamma=0$. If we set $\alpha=\beta=\gamma=0$ than our previous analysis is complete for the commutation relations of $A$.  If we go with the $a=0$ option, this implies that the $\mu_1 R$  term cannot be included, since as we showed above, for $a=0$ and $\mu_1\neq 0$, then $b=0$ and $c=0$ (and one of our conditions was exactly that we need to include either $a, b,c \neq 0$).
From this point, it will be easy to see that the commutation relations of the 
$\alpha, \beta, \gamma$ terms with the $b$ and $c$ terms  necessarily produce terms
 proportional to $R$, and thus cannot be weakly zero, unless the coefficients explicitly cancel (in which case they are strongly zero).

 So let us get to it. We start, once again for simplicity, with a trial term, the term $(\alpha, a)$:
\begin{equation}
\alpha a:~~ \int d^3 x N_2\left(-\frac{1}{2}(\nabla^2(R^{ab}N_1)+g^{ab}(NR^{cd})_{;cd}-2((NR^{c(b})_{;c})^{;a)}\right)\pi_{ab} - (N_1\leftrightarrow N_2)
\end{equation} 
 There are some terms that we will discard at a first brush. Namely, any of the terms proportional to the trace of the momenta (because we will pessimistically assume that they might be canceled by the $(\alpha, b)$ contributions, as it happened with the $R^n$ terms and $a,b$), and those that have the outermost covariant derivative contracted with the momenta. In this case we are left only with:
 \begin{equation}\label{equ:simplified_bracket}
(\alpha, a):~~ \int d^3 x N_2\left(-\frac{1}{2}(\nabla^2(R^{ab}N_1)\right)\pi_{ab} - (N_1\leftrightarrow N_2)
\end{equation} 

 Now we write the general $(\alpha,\beta,\gamma, a)$ term (to simplify notation we will denote $N_1\rightarrow N$)
 \begin{eqnarray}
(\alpha,\beta,\gamma, a): ~~ a\int d^3 x N_2\Big(-\frac{\alpha}{2}\left(\nabla^2(R^{ab}N)+g^{ab}(NR^{cd})_{;cd}-2((NR^{c(b})_{;c})^{;a)}\right)\nonumber \\
-\frac{\beta}{2}\Big(\nabla^2(N^{;ab})+(N^{;cd})_{;cd}g^{ab}-(N^{;cba})_{;c}-(N^{;bca})_{;c}
+2((NR^{c(a})_{;c})^{;b)}-(NR^{cd})_{;cd}g^{ab}\nonumber\\
+(\nabla^2(R^{ab}N)-4(R^{c(a}_{~~;c}N)^{;b)}+2(R^{cd}_{~~;d}N)_{;c}g^{ab}
-(R^{c(a;b)}N)_{;c}+(R^{ab;c}N)_{;c}\Big)\nonumber
\\
+\gamma\Big(-\nabla^2(\nabla^2N)g^{ab}+(\nabla^2N)^{;ab}-\nabla^2(R^{ab}N)+2(NR^{ab;c})_{;c}
+(NR^{;a})^{;b}-\frac{1}{2}(NR^{;c})_{;c}g^{ab}\Big)
\Big)\pi_{ab}=0\label{equ:full_variation}
\end{eqnarray}

After the same simplifications as applied in \eqref{equ:simplified_bracket},  the result is given by:
\begin{eqnarray}
(\alpha,\beta,\gamma, a)~&:&a\int d^3 x N_2\Big(-\frac{\alpha}{2}\left(\nabla^2(R^{ab}N)\right)\nonumber \\
&~&~-\frac{\beta}{2}\left(\nabla^2(N^{;ab})-(N^{;cba})_{;c}-(N^{;bca})_{;c}+(\nabla^2(R^{ab}N)-2(R^{c(a;b)}N)_{;c}+
(R^{ab;c}N)_{;c}\right)\nonumber\\
&~&~+\gamma\left(-\nabla^2(R^{ab}N)+2(NR^{ab;c})_{;c}\right)
\Big)\pi_{ab}=0\label{equ:simple_full}
\end{eqnarray}
Our task is to show that the only solution to this equation for the constants $(\alpha, \beta, \gamma, a)$ is either $(\alpha, \beta, \gamma, 0)$ or $(0,0,0,a)$, as we mentioned. That is, our task is to show that the three lines above are linearly independent, which seems very intuitive, since for example, the $\beta$ line has a different structure of free indices, which can contract with an arbitrary symmetric tensor $\pi_{ab}$. 

The easiest way to show this explicitly is to see that in the second line of \eqref{equ:simple_full} (the $\beta$ line) there are terms which do not include the Riemmann curvature (e.g. don't vanish if the Riemann curvature vanishes), and thus cannot be canceled by the other terms. Then, left with the $\alpha$ and $\gamma$ it is trivial to see that they can't cancel. 
To be more precise, the term $(N^{;cba})_{;c}\pi_{ab}$ does not include the Riemann curvature, which would be  anti-symmetric for the indices $a, b$. 
In other words,  the four-derivative terms of $N$ are
  $$\nabla^2(N^{;ab})-2(N^{;bca})_{;c} =-\nabla^2(N^{;ab})+{R^{bca}}_{d;c}N^{,d}$$ which means that   we are always left with pure 4-th derivative of $N$ terms which do not  depend on the Riemann curvature, and thus $\beta=0$, which means that $\alpha=\gamma=0$ as well. Thus let us move on to the analysis of the consequences of $a=\mu_2=\mu_1=0$ (and $\alpha, \beta, \gamma$  arbitrary).

  Unlike the contraction with $\pi_{ab}$  on \eqref{equ:full_variation}, we will now have the contraction with either $g_{ab}$  or $\pi g_{ab}$. This will involve taking the trace of the expression being contracted with $\pi_{ab}$ in \eqref{equ:full_variation}. Using the contracted Bianchi identity $R^{ab}_{~~;b}=\frac{1}{2}R^{;a}$, the only terms that
  can arise  from \eqref{equ:full_variation} (with contraction with $(b\pi+c) g_{ab}$ instead of $\pi_{ab}$) are of the form $N^{;cd}R_{cd} ,~ \nabla^2\nabla^2N, ~N\nabla^2R, ~R\nabla^2N$ and $R_{;c}N^{;c}$, which forms the linearly independent basis of the terms whose coefficient will have to vanish. \footnote{One cannot do integration by parts to cancel between the terms $R_{c}N^{;c},~R\nabla^2N$  and $N\nabla^2 R$ (since everything is multiplied by $N_2$ or $N_2\pi$).For example, suppose you had the combination $aR_{;c}N^{;c}+bR\nabla^2N+cN\nabla^2 R$. Upon various integration by parts to isolate $N$ for instance, one gets the conditions $a-b+c=2a-b=c=0$, whose solution is only $a=b=c=0$. }  Terms such as $N_{;ac}N^{;ac}$ can be absorbed into the rest, as follows: 
   $N^{;bca}_{~~~~;c}=N^{;cba}_{~~~~;c}$ and  $N^{;bca}_{~~~~;c}g_{ba}=N^{;cd}_{~~;cd}$, thus 
\begin{equation}\label{4thN}g_{ab}(N^{;bca})_{;c}=g_{ab}(R^{bca}_{~~~d}N^{;d})_{;c}+N^{;bac})_{;c})=(R^{cd}N_{;c})_{;d}+\nabla^2\nabla^2N
   \end{equation}

 We start by considering the  terms $\nabla^2\nabla^2N$ and $R_{cd}N^{;cd}$ (these include, as shown above, the linearly dependent term $N^{;cd}_{~~~;cd}$). We will find that the vanishing of the coefficients of these terms will demand that $\alpha=0$ and $\beta=-2\gamma$. 
let us start by rewriting the appropriate part of the  $\beta$ term from  \eqref{4thN}
\begin{equation}
\label{equ:beta_4thN} (\nabla^2(N^{;ab})+(N^{;cd})_{;cd}g^{ab}-(N^{;cba})_{;c}-(N^{;bca})_{;c})g_{ab}
=2\nabla^2\nabla^2N+(R^{cd}N_{;c})_{;d}
\end{equation}
 Using \eqref{equ:beta_4thN} and \eqref{equ:full_variation}, we get the following coefficients 
 $$ -\frac{\alpha}{2}R_{cd}N^{;cd} ~~\mbox{and}~~(-\beta-2\gamma)\nabla^2\nabla^2 N
 $$
 which implies the promised  $\alpha=0$ and $\beta=-2\gamma$.\footnote{It is very easy to see that these terms are linearly independent from each other and from the rest: the $R^{ab}$ term doesn't necessarily vanish even if $R$ vanishes, and $\nabla^2\nabla^2 N$ doesn't vanish even if $R^{ab}$ vanishes. Similarly it is not hard to show that all terms of our basis are in fact linearly independent.} 
 
 At this point, it is worth stressing once again that since $\mu_2=\mu_1=0$, if we integrate by parts the contracted derivatives, we have to obtain a vanishing coefficient for $R$ in order that the $A$  have any chance of weakly commuting with itself.
 Let us continue our attempt to find the conditions under which each coefficient of each term vanishes. It turns out that the coefficient of the term $R\nabla^2N$ does not yield any new conditions apart from $\alpha=0$ and $\beta=-2\gamma$, but the coefficient of $N\nabla^2R $ is given by 
  $$\frac{3\alpha}{4}-\beta-\frac{\gamma}{2}=0$$
  which together with our previous conditions demand that $\beta=\gamma=0$. We have thus proven:
  \begin{theo}\label{theo:prop}
  Given the constraints $A=0$ and $\pi^{ab}_{~~;a}=0$, where
  $$A(\alpha,\beta,\gamma,\mu_n,a,b,c):= \alpha R^{ab}R_{ab}+\beta \nabla_a\nabla_bR^{ab}+ \gamma\nabla^2R + \mu_n R^n+\frac{a\pi^{ab}\pi_{ab}+b\pi^2}{g}+ c\frac{\pi}{\sqrt g}$$
and we use the summation rule for $n$, the only choice of coefficients which have at least one of $a,b,c\neq 0$ for which $A$ weakly commutes with itself, are $(\alpha,\beta,\gamma,\mu_n,a,b,c)=(0,0,0,\mu_1,-2b,b,0)$ and $ (0,0,0,0,a,b,c)$.
  \end{theo}
We should mention that some of these same calculations (with less generality) have been performed in \cite{RWR}, obtaining results agreeing with our own. 
 
\subsection{Gauge-fixing (second-class) properties of the constraints} 
 
 The second part of the proof of our results goes as follows. We have the following set of constraints (where we have eliminated the $\alpha, \beta,\gamma$ components altogether): $(\mu_n,a,b,c)=\{(\mu_1,-2b,b,0), (0,a,b,c)~|~\mu_1, a,b,c \in \mathbb{R} \}$, each of which forms a first class system when taken together with the diffeomorphism constraint. The way that we cut down further on these is by exploring the conditions required for symmetry doubling. In other words, we will prefer any pair of such constraints that gauge-fix each other generically in phase space. \footnote{In fact, although genericity in the technical sense can easily be shown (see footnote \ref{footnote:generic}) , this is not what we do in the main text. We demand that the set in which the constraints \emph{fail} to gauge-fix each other cannot be infinite-dimensional. Failure to gauge-fix  at a given point in phase space $(g,\pi)$, here means that the kernel of the Dirac matrix is not necessarily finite-dimensional. } To illustrate the power of these further restrictions independently of what has already been cut down by the previous principle (section \ref{sec:first_class}), we will  use the more general constraints set: 
 $$(\mu_n,a,b,c)=\{(\mu_n,-2b,b,0), (0,a,b,c)~|~\mu_n, a,b,c \in \mathbb{R} \}$$
 thus allowing in the calculations terms like $R^n$.

\subsubsection{Cross-terms: $(\mu_n,-2b,b,0)$ and $(0,a,b',c)$.} 
  We will start the calculation between the possible pairs  $(\mu_n,-2b,b,0)$ and $(0,a,b',c)$. We will then perform the calculation for the pairs $(\mu_n,-2b,b,0)$, $(\mu_n',-2b',b',0)$ and finally $(0,a,b,c)$, $(0,a',b',c')$.

 The general Poisson bracket for our starting term is:
 \begin{multline}
 \{(\mu_n,-2b,b,0)[N],(0,a,b',c)(x)\}:=\\
 \{\int d^3x'\sqrt gN\left( \mu_n R^n+\frac{-2b\pi^{ab}\pi_{ab}+b\pi^2}{g}\right)(x'), \left(\frac{a\pi^{ab}\pi_{ab}+b'\pi^2}{\sqrt g}+ c{\pi}\right)(x)\}
 \end{multline}
 where just as a reminder, we are using the summation rule for $n$. The difference between the present calculation and the one done previously (to find which constraints commuted with themselves), is that now we are looking at two different sets of constraints, which come with different smearings, and thus terms linear in the smearings also contribute, which makes the calculation slightly more involved. But we have already done the hard work. 

 The smeared variation of the local $(\mu_n,-2b,b,0)$ is:
\begin{multline}\label{equ:delta_gSN}\diby{\int d^3x' N(x')(\mu_n,-2b,b,0) (x')}{ g_{ef}(y)}=
\left(\frac{b}{\sqrt g} g^{ef}G_{abcd}\pi^{ab}\pi^{cd}-\frac{4b}{\sqrt g}(\pi^{eb}g_{bd}\pi^{fd}-\frac{\pi^{ef}\pi}{2})\right)N(y)\\
-\mu_n\Big(\frac{1}{2}\sqrt g g^{ef}R^nN(y)+
n\sqrt g(y)(-R^{n-1}R^{ef}(y)-g^{ef}(y)\nabla^2(R^{n-1}N)(y)+
(R^{n-1}N)^{;ef}(y) )\Big)
\end{multline}
where $G_{abcd}=g_{ac}g_{bd}-\frac{1}{2} g_{ab}g_{cd}$ is the inverse DeWitt metric. 

Now for the momentum variation:
\be\label{equ:delta_piS} \diby{\int d^3x' N(x')(\mu_n,-2b,b,0) (x)}{ \pi^{ef}(y)}=
-4b\frac{G_{efcd}\pi^{cd}}{\sqrt g}N(y)=-2b\frac{2\pi_{ef}-g_{ef}\pi}{\sqrt g}(x')N(y)
\ee
For our second set: 
\be\label{equ:delta_piSD}\frac{\delta (0,a,b',c)(x)}{\delta \pi^{ef}(y)}=\left(2\frac{a\pi_{ef}+b'g_{ef}\pi}{\sqrt g}(x)+cg_{ef}\right)\delta(x,y)
\ee
and 
 \be\label{equ:delta_gSD}\frac{\delta (0,a,b',c)(x)}{\delta g_{ef}(y)}=\left(
 -\frac{g^{ef}}{2\sqrt g} (a\pi^{ab}\pi_{ab}+b'\pi^2)+\frac{2}{\sqrt g}(a\pi^{eb}g_{bd}\pi^{fd}+b'{\pi^{ef}\pi})+c\pi^{ef}\right)\delta(x,y)
\ee

We now integrate \eqref{equ:delta_gSN}$\cdot$\eqref{equ:delta_piSD}$-$ \eqref{equ:delta_piS}$\cdot$\eqref{equ:delta_gSD} over $y$. It turns out that the coefficient of the contracted product of three momenta $\pi^a_c\pi^c_b\pi^b_a$ vanishes. The final result is:
\begin{multline}\label{equ:DeltaN}
 \{(\mu_n,-2b,b,0)[N],(0,a,b',c)(x)\}:=\Delta N\\
=N\left( (2bb'+ab)\pi\frac{3\pi_{ab}\pi^{ab}-\pi^2}{g}-3bc\frac{G_{abcd}\pi^{ab}\pi^{cd}}{\sqrt{g}}\right)\\
-n\mu_n\sqrt{g}\left(\frac{N}{n}R^n(\frac{\pi}{\sqrt{g}}(b'+a)+\frac{c}{2})+\nabla^2(R^{n-1}N)(-2\frac{\pi}{\sqrt{g}}(2b'+a)-2c)+2a(-NR^{n-1}R^{ef}+(R^{n-1}N)^{;ef})\pi_{ef}\right)
  \end{multline}
where we have defined the phase space dependent  second order differential operator $\Delta$ acting on the smearing function $N$. We would like to make choices for the coefficients $a,b,b',c,\mu_n$ such that the operator is  most generically invertible.  
Invertibility is equivalent, in the closed manifold case, for there being no non-zero homogeneous solution, i.e.  $\Delta u=0$ implies $u=0$. 

The terms with the highest order derivatives acting on $N$ are:
\begin{equation}
\label{equ:highest_order}
n\mu_n\left(R^{n-1}\nabla^2N(-2\frac{\pi}{\sqrt{g}}(2b'+a)-2c)+2aR^{n-1}N^{;ef}\pi_{ef}\right)
\end{equation}
These terms define the principal symbol of the differential operator, which in turn determines whether the equation can be characterized as elliptic, parabolic or hyperbolic. If the principal symbol of $\Delta$ is itself invertible (has no zeroes), the operator is called \emph{elliptic}. Ellipticity itself already guarantees that the kernel of $\Delta$ is finite-dimensional.  The zeroes of the principal symbol correspond to the characteristics of either a parabolic  (i.e. a heat equation) or a hyperbolic equation (a wave equation). Either of these two types of equations have very large sets of zero functions, e.g. $\Delta u=0$ have many solutions where $u\neq 0$, since in these cases $\Delta u=0$  can be solved by an infinite amount of ,e.g. wave functions $u$ which depend on some initial data. Thus our first criterion will be to investigate the symbol of the operator $\Delta$, eliminating those choices of coefficients for which it is not generically elliptic.  

The principal symbol $\sigma(D)$ of a differential operator $D$ is intimately tied to a Fourier transform: it involves replacing the derivatives by an element of the cotangent bundle (the momenta), $\partial_a\rightarrow p_a$. The symbol will be invertible, and thus the operator elliptic, if the only value for which it vanishes is  $p_a=0$.  We thus write, from \eqref{equ:highest_order}: 
\begin{equation}
\label{equ:principal_symbol}
\sigma_p(\Delta)=\mu_nR^{n-1}((-2\frac{\pi}{\sqrt{g}}(2b'+a)-2c)p_\alpha p^\alpha +2a\pi^{ef} p_ep_f)
\end{equation}
First, we check that  $\sigma$ loses ellipticity for infinite-dimensional sets of phase space for $\mu_{n}\neq 0$,  $n>1$. The argument below mainly aims to show that not too much fine-tuning is needed to find metrics which have zero curvature in at least some open set of the spatial manifold $M$. 

For any $n>1$, setting $R=0$ (even locally) the operator $\sigma_p(\Delta)$ is no longer elliptic. Given any initial metric $g$ not in the positive Yamabe class, and an open set $O$ of the spatial manifold $M$, one can find a conformal transformation of the metric that brings it to a metric of $R[\tilde g]_{|{O}}=0$. This is easy to see as follows: suppose that the metric is in the negative Yamabe class (since if it were in the zero Yamabe class we could trivially set $O=M$). Then one can find a conformal representative $\tilde g$ of $g$ such that $R[\tilde g]=- c^2$. Performing a further conformal transformation we obtain $R[\tilde{\tilde g}]=\nabla^2\Omega+ c^2\Omega$. Since we are only looking at an open set (or with arbitrary auxiliary boundary conditions for $\Omega$), the spectrum of the Laplacian is negative and continuous, which allows us to set $R[\tilde{\tilde g}]_{|O}=0$. This is enough for our claims of genericity (note as well that the momenta is still completely unspecified) although it is easy  to extend the argument above  for positive Yamabe class. \footnote{As mentioned in the beginning of the section, we have shown this for open infinite-dimensional subsets of phase space. All one needs to do to show this generically in phase space (in the technical sense) is, given an initial metric $g$, to find an arbitrarily small deformation $\delta g$ of $g$ (small in e.g. the Whitney or $C^\infty$ topology) such that $g+\delta g$ has zero scalar curvature on some small open set $O$. By making the set small enough, one may perturb the metric arbitrarily so that one achieves the zero scalar curvature in some open set, while still keeping the metric $g+\delta g$ arbitrarily close to $g$ (in the given topology of configuration space).  To be slightly more specific, since topologically $\mathbb{R}^n$ admits metrics of arbitrary negative scalar curvature, using a bump function one can easily find a small set $O\subset M$ for which $\delta g$ ``evens out" the curvature there. Although this statement is quite obvious physically, we will not attempt to prove it formally here, as it involves too many further notions ($C^\infty$ topology, bump functions, curvature of composite metrics, etc). \label{footnote:generic}}

  We thus set  $\mu_{n}= 0$ if $n>1$, and are left only with $\mu_1=:\mu$. 
The ellipticity of \eqref{equ:principal_symbol} for this choice thus requires that $\mu\neq 0$ and  
\be \label{equ:princ_2}
(-2\frac{\pi}{\sqrt{g}}(2b'+a)-2c)p_\alpha p^\alpha +2a\pi^{ef} p_ep_f\neq 0
\ee
Clearly there are large sets of choices of $\pi^{ab}$  for which the above vanishes at some points in space, given any choice of coefficients. 

For example, we can diagonalize the real symmetric tensor $\pi^{ab}$ (by orthogonal matrices which thus leave $g_{ab}$ unchanged), with given eigenvalues $\lambda_1,\lambda_2,\lambda_3$. Let us consider then that the metric in some neighborhood as  given by the diagonal Euclidean metric,  $g_{ab}=\eta_{ab}$.\footnote{As will become clear in the following, the method we utilize would only show more violations of ellipticity had we considered the general diagonal metric. Although complexity rises minimally, we felt that the extra complication was unnecessary to prove our point. } We get from  \eqref{equ:princ_2}, that the following three quantities should be non-zero:
\be \label{equ:zeroset}
(-2(\lambda_1+\lambda_2+\lambda_3)(2b'+a)-2c) +2a\lambda_i\neq 0
\ee
since if the $i$th equation is zero, the symbol is clearly not injective (all the terms with $p_i\neq 0$ and $p_j=0$ for $j\neq i$ are in the kernel). 

But it is easy to find choices for which the lhs of \eqref{equ:zeroset} vanishes,  for example for $i=1$, if $b'\neq 0$ 
$$\lambda_1=\frac{(\lambda_2+\lambda_3)(2b'+a)-2c)}{2b'}
$$
does not obey \eqref{equ:zeroset}. Thus for a large 2 parameter set of functions, we have non-elliptic symbols. If $b'=0, a\neq 0$, for $i=1$ we get also a two-parameter set of solutions $\lambda_2=\frac{c}{a}-\lambda_3$ and $\lambda_1$ unrestricted. One can easily adapt the argument  for merely a diagonal metric $g_{ab}=\mbox{diag}(\kappa_1,\kappa_2,\kappa_3)$, obtaining even larger sets of solutions. This implies that for either of these choices the operator would not be generically elliptic. 

Thus we are left with $a=b'=0$, for which  we get from \eqref{equ:princ_2}:
$$ cp_\alpha p^\alpha\neq 0
$$
which holds whenever $c\neq 0$.

With these restrictions, \eqref{equ:DeltaN} becomes: 
\begin{equation}
\Delta N=c\left(
N\left( -3b\frac{G_{abcd}\pi^{ab}\pi^{cd}}{\sqrt{g}}\right)
-\mu\sqrt{g}\left(\frac{1}{2}NR-2\nabla^2N\right)\right)
  \end{equation}
which is elliptic, and generically will have only the trivial kernel, but still might have a non-zero \emph{finite-dimensional} kernel at some singular subsets of phase space.

\subsubsection*{Ellipticity restricted to the the intersection surface}

In some sense, we can further the claim of theorem \ref{theo:gauge-fix1} by using the weak equalities of the constraints. I.e. we can analyze the ellipticity of the operator on the intersection of the constraint surfaces itself.  
By weakly we mean that we can set the combinations 
\be\label{equ:weak_equ}\mu_n R^n+\frac{-2b\pi^{ab}\pi_{ab}+b\pi^2}{g}\approx 0\mbox{~~and~~} \frac{a\pi^{ab}\pi_{ab}+b'\pi^2}{g}+ c\frac{\pi}{\sqrt g}\approx 0\ee
 For instance, {it is easy to see that  if we used the weak equalities the statement leading to the setting of $\mu_n=0$ (except for $\mu_1$) implies only that $\frac{-2b\pi^{ab}\pi_{ab}+b\pi^2}{g}=0$, which is still quite a large set on phase space, since this equation only determines one degree of freedom of the momenta. } 
 
  Similarly, the other restriction $b'=c=0$ would still hold, but one must use the full diagonal metric $g_{ab}=\mbox{diag}(\kappa_1,\kappa_2,\kappa_3)$ in the ansatz leading to theorem \ref{theo:gauge-fix1}: one of the $\kappa$ functions can be used to solve the first weak equality of \eqref{equ:weak_equ}, and one of the $\lambda$'s to solve the second. One can show that there would still be left a 3-parameter family of solutions for these cases. This is so because we would only need to have one of the analogous  three inequalities \eqref{equ:zeroset} broken. Thus we would have 6 free functions for 3 equations, whereas in the present simplified case we had 3 free functions for one equation. 
  
   We can now use the weak equalities \eqref{equ:weak_equ} to investigate the  kernel of $\Delta$ given in \eqref{equ:DeltaN} on the intersection of the constraint surfaces. We have the equality: 
$$\mu R\approx 2b \frac{G_{abcd}\pi^{ab}\pi^{cd}}{{g}}\approx 2b \pi^{ab}\pi_{ab}$$
which yields: 
\begin{equation}\label{equ:DeltaN_approx}
\Delta N\approx 2c
\left(\nabla^2N-2b \frac{\pi^{ab}\pi_{ab}}{g}\right)N
  \end{equation}
Now according to the min-max principle,  since $\pi^{ab}\pi_{ab}>0$, as long as $b> 0$, we have no non-zero homogeneous solution on a closed manifold.\footnote{The min-max principle is quite simple: suppose we are on a closed  manifold, then if $N>0$ there exists a point $x_o$ for which $N(x_o)$ is maximum, which implies that $\nabla^2N(x_o)<0$, for which there can be no solution. The same holds for $N<0$ and the minimum.  } Of course, even if $b<0$ we might generically not have any kernel, but we do not know how to classify the kernel of $\Delta$ for $b<0$. 

There is a new issue at the constraint surface which is determining where the constraint surfaces intersect. For example,  $b>0$ implies that the scalar curvature has the same sign as the coefficient $\mu$, whereas $b<0$ it must have opposite sign. These limitations can be overcome by not setting the constraints to zero, but to a given constant (e.g. a cosmological constant in the case of the scalar curvature and a constant trace of the momenta instead of a zero trace). \footnote{This is what is done in Shape Dynamics \cite{SD_first}, which in the end allows one to have the gauge fixing and the intersection of the constraint surfaces not limit the sign of the scalar curvature. The treatment of this case is beyond the scope of this paper. }
 
 \subsubsection*{The remaining cases:  $\{(\mu_n,-2b,b,0)$, $(\mu'_{n'},-2b',b',0)\}$ and $\{(0,a,b,c)$, $(0,a',b',c')\}$}
  These calculations are made much easier still by the work of the previous section. 

 For the highest order derivative coming from the $\{(\mu_n,-2b,b,0)$, $(\mu_{n'}',-2b',b',0)\}$ term, we obtain, as in \eqref{equ:highest_order}:
 \be\label{equ:diagonalPB}
 \left(4N^{;ef}\pi_{ef}\right)(R^{n-1}n\mu_nb'-R^{n'-1}n'\mu_{n'}'b)
\ee
since in this case the quantities $2b'+a$ and $c$ of \eqref{equ:highest_order} vanish. From the previous arguments it is easy to see that this only has a chance at being generically second class if 
 $(R^{n-1}n\mu_nb'-R^{n'-1}n'\mu_{n'}'b)=0$, which implies that $n=n'$ for all $n $ and $n'$, and thus it is not hard to see that we must have $\mu_nb'=\mu_{n}'b$. We can absorb $b$ and $b'$ into $\mu_n$ and $\mu_n'$, which implies $\mu_n=\mu_n'$ which implies  that we have two copies of the same constraint (and thus first class by the results of the previous section). 
 
 For the final case $\{(0,1,a,b)$, $(0,1,a',b')\}$, where we have absorbed an overall constant in each constraint, assuming the original $a\neq 0$. We obtain 
\be\label{equ:pipi_bracket} \frac{3}{g}\left((\pi^{cd}\pi_{cd}(\frac{\pi}{\sqrt g}(a-a')-\frac{b-b'}{2})+\frac{\pi^2}{6}(2(a-a')\pi+3(ab'-ba'))\right)
\ee
 which has an infinite-dimensional kernel in the infinite-dimensional subspace $\pi^{ab}=0$. \footnote{This would be enough for our purposes, but it does not possess the same level of genericity as the previous analysis.  We can go further: substituting $\frac{\pi^{ab}\pi_{ab}}{g}$ of one constraint into the other we obtain: 
 $ (a'-a)\pi^2+(b'-b)\pi\sqrt{g}=0 .
 $ If $b\neq b'$ the solutions are $\pi=0$, $\pi=-\frac{b'-b}{a'-a}$. The $\pi=0$ solution implies that $\pi^{ab}=0$ through either of the constraints, if $a'=a$  we also have $b=b'$. Provided $a\neq a, b\neq b'$ and $\pi\neq 0$, the second solution, $\pi=-\frac{b'-b}{a'-a}$, inserted back into \eqref{equ:pipi_bracket} is the only one that could still provide conditions on the coefficients to form an invertible bracket at the constraint surface. After some algebra we obtain the extra conditions on the coefficients such that this also does not vanish on the intersection surface: $b(a-a')-\frac{2}{3}(b-b')\neq 0$. } 
 
 We have thus proven:
\begin{theo}\label{theo:gauge-fix1}
Given the two sets of constraints $\{A_i=0$ and $\pi^{ab}_{~~;a}=0\}_{i=1,2}$ , where
  $$A_i(\alpha_i,\beta_i,\gamma_i,\mu_n^{(i)},a_i,b_i,c_i):= \alpha_i R^{ab}R_{ab}+\beta_i \nabla_a\nabla_bR^{ab}+ \gamma_i\nabla^2R + \mu_n^{(i)} R^n+\frac{a_i\pi^{ab}\pi_{ab}+b_i\pi^2}{g}+ c_i\frac{\pi}{\sqrt g}$$
the only choices of coefficients which: i) have at least one of $a,b,c\neq 0$, ii) for which $A_i$ weakly commutes with itself, and iii) the commutator between  $A_1$ and  $A_2$ has (generically) at most a finite-dimensional kernel  are:
 $(\alpha,\beta,\gamma,\mu_n,\mu_{n-1},\cdots,\mu_1, a,b,c)=(0,\cdots,0,\mu_1,-2b,b,0)$ and $ (0,\cdots,0,c)$. \end{theo}
\subsection{Summary of results of this section} 
 
\paragraph*{The results}
Here we  present a summary of the results of the calculations. The way we performed the calculations 
is simple. The terms that we  consider are (remembering that at least one of the roman letters cannot be equal to zero):
 $$A(\alpha,\beta,\gamma,\mu_n,a,b,c):= \alpha R^{ab}R_{ab}+\beta \nabla_a\nabla_bR^{ab}+ \gamma\nabla^2R + \mu_n R^n+\frac{a\pi^{ab}\pi_{ab}+b\pi^2}{g}+ c\frac{\pi}{\sqrt g}$$
and we use the summation rule for $n$.  We first calculated the Poisson bracket
$ \{A(x),A(y)\}$ and demanded that this Poisson bracket vanish weakly (i.e. when simultaneously $\pi^{ab}_{;b}=0$  (the spatial diffeomorphism constraint)  and $A(x)=0$). 
The result of this part of the calculation is theorem \ref{theo:prop}, which drastically limits the possible constraints to  $(\alpha,\beta,\gamma,\mu_n,a,b,c)=(0,0,0,\mu_n,-2b,b,0)$ and $ (0,0,0,0,a,b,c)$.  

Among these candidates we looked for sets that would also generically gauge-fix each other, thereby producing a symmetry doubling pair \cite{SD:Sym_doub}. To do so, we demanded that the Poisson bracket between them form an invertible operator, a condition which can be rephrased in more field theoretic language as the demand that the Fadeev-Popov determinant in phase space be non-zero. The main theoretical tool to investigate this is to look for ellipticity of the operator, as this will imply that the operator have at most a finite-dimensional kernel, whereas failure of ellipticity will imply that the kernel is infinite-dimensional\footnote{The operator becomes a wave or heat operator, with generically well-defined initial value problem.}  This criterion of ellipticity simplifies the search considerably, and yields theorem \ref{theo:gauge-fix1}. 

Thus we are left with the only two possibilities: $(\alpha,\beta,\gamma,\mu_n,\mu_{n-1},\cdots,\mu_1, a,b,c)=(0,\cdots,0,\mu_1,-2b,b,0)$ and $ (0,\cdots,0,c)$. The latter is just the maximal slicing constraint $\pi\simeq 0$, while the former is a one-parameter family of constraints, which we now parametrize by $\lambda$, 
$$ \lambda R - \frac{\pi^{ab}\pi_{ab}-\frac{1}{2}\pi^2}{g}\simeq 0
$$
which includes the usual ADM scalar constraint for $\lambda=1$.

\section{Conclusions}\label{sec:conclusions}

\subsubsection*{A preferred slicing for ADM}

Although maximal slicing is a very popular gauge  in numerical studies of general relativity, it is not considered to be fundamental in any sense. Here we have shown  that there is a very precise sense in which this gauge choice is  special, and it is  not  because it is computationally convenient. It emerges naturally from symmetry principles, \emph{ simultaneously} with a slight generalization of ADM. Although the respective Hamiltonian phase-space reduced theory is not identical to the full fledged theory of general relativity, it has a very large intersection in terms of observables, and serves to disentangle a feasible theory of gravity from the concept of time inherent to general relativity.

\subsubsection*{Independence from GR}

The criticism of Shape Dynamics dealt with in this paper is that the standard construction of the theory depends heavily on the existence of ADM gravity. Indeed, as mentioned in the introduction, Shape Dynamics is obtained by introducing a Stuckelberg field - implementing spatial Weyl invariance - and subsequently gauge-fixing the extended theory in differing ways.  Here we have shown and motivated a set of axioms that simultaneously generate both the ADM type of refoliation symmetries (a one parameter family which includes the ADM Hamiltonian) and the Weyl symmetry generator. The reduced reduced phase space theory obtained from the imposition of both sets of constraints does not have as canonical kinematical variables $g_{ab}$ and $\pi^{ab}$ - it cannot be said to be geometrical (as reduced phase space theories in general). But two distinct natural  extensions exist that have exactly the set $g_{ab}$ and $\pi^{ab}$ as canonical kinematical variables: ADM and Shape Dynamics.  In our opinion this is a large step in the direction of independence of Shape Dynamics from ADM, or, if not in the direction of independence, in that of a an equality between Shape Dynamics and ADM gravity.

 
\subsubsection*{Relation to exact renormalization group arguments for gravity.}

  A  working hypothesis for the exact renormalization group arguments for quantum gravity is that  although gravity does not possess a Gaussian fixed point, it might possess a non-Gaussian one. In that case, the different symmetry content of Shape Dynamics might be more appropriate for the search of such fixed points.
As has been argued before by Koslowski and Eichhorn \cite{Astrid}, \cite{Tim_effective}, symmetry doubling could offer an interesting new tool in the search for an asymptotically safe Quantum gravity. In the present work we have demonstrated that the emergence of the two sets of  symmetries generating refoliations and the Shape Dynamics spatial Weyl symmetry, follows almost straightforwardly from the requirements for symmetry doubling.\footnote{In \cite{SD:Sym_doub} it was shown that the necessary condition for symmetry doubling to occur is that the phase space possess two sets of mutually gauge-fixing sets of first class constraints.}
One spin-off  should be  to  put Shape Dynamics to use in the exact renormalization group and asymptotic safety scenario. First steps in that direction would be to extend our results to include a cosmological constant in the definition of the functional $A$ in \eqref{def:A}, and to reframe Reuter's approach \cite{Reuter} in the Hamiltonian setting. This extension could already provide an interesting 2-dimensional truncation of terms to be considered as an input in (an appropriate Hamiltonian translation of) the machinery developed by Reuter \cite{Reuter}. 


\subsubsection*{The issue with the solution.}

There are three issues we should call attention to in our ``solution". First and foremost, we have used an asymptotic safety argument as a starting point heuristic for our requirements on the constraints \cite{Astrid, Tim_effective}. We have no proof that this heuristic can be turned into a formal proof for such requirements.  Secondly, we have only performed the calculations up to 4th order in spatial derivatives. Arguments for why one can extend this beyond this order have been given by Koslowski in personal communication, and a follow up in this direction is in preparation. They rely on the fact that higher order derivative terms would not disturb the universality class of the model.  Thirdly, we have two ``extra" assumptions that might be objectionable to: the inclusion of ``time" in the constraints - by demanding that they include momenta - and the exclusion of terms that have derivative coupling. The first limitation we feel is fundamental, and might be further justified by making an analogy between ``time" and ``renormalization time" \cite{Strominger}. Namely,  if we do include the possibility of such terms that are intrinsic to the geometry of the hypersurface they might form the analogues of conformal fixed points, for trajectories that already sit at the fixed points, and thus where nothing ever ``happens". The second assumption is of technical value, but we feel that the final result of the calculation might not suffer from its removal, although it becomes much more complicated.  Furthermore a working  assumption of excluding derivative couplings in gravity is quite common in many different circumstances. 

\section*{Acknowledgements}

The author would like to thank  Tim Koslowski for ongoing conversations on this topic. I would also like to thank Steve Carlip for reading the manuscript and suggesting several important improvements. HG was supported in part by the U.S.
Department of Energy under grant DE-FG02-91ER40674.


\begin{thebibliography}{10}

\bibitem{Nima-Lectures}
N.~Arkani-Hamed, ``Space-time, quantum mechanics, and the large hadron
  collider.'' IAS Lecture. http://video.ias.edu/space-time.

\bibitem{SD:Birkhoff}
H.~Gomes, ``{A Birkhoff theorem for Shape Dynamics},'' {\em ARXIV:1305.0310},
  2013.

\bibitem{Lovelock}
D.~{Lovelock}, ``{The Four-Dimensionality of Space and the Einstein Tensor},''
  {\em Journal of Mathematical Physics}, vol.~13, pp.~874--876, June 1972.

\bibitem{HKT}
S.~A. Hojman, K.~Kuchar, and C.~Teitelboim, ``{Geometrodynamics Regained},''
  {\em Annals Phys.}, vol.~96, pp.~88--135, 1976.

\bibitem{SD_first}
H.~Gomes, S.~Gryb, and T.~Koslowski, ``{Einstein gravity as a 3D conformally
  invariant theory},'' {\em Class. Quant. Grav.}, vol.~28, p.~045005, 2011.

\bibitem{SD:Sym_doub}
H.~Gomes and T.~Koslowski, ``{Symmetry Doubling: Doubly General Relativity},''
  {\em ARXIV:1206.4823}, 2012.

\bibitem{Reuter}
M.~Reuter, ``{Nonperturbative evolution equation for quantum gravity},'' {\em
  Phys.Rev.}, vol.~D57, pp.~971--985, 1998.

\bibitem{Tim_effective}
T.~A. Koslowski, ``{Shape Dynamics and Effective Field Theory},'' {\em
  Int.J.Mod.Phys.}, vol.~A28, p.~1330017, 2013.
  
  
  \bibitem{RWR}
J.~Barbour, B.~Foster, and N.~O. Murchadha, ``Relativity without relativity,''
  {\em Classical and Quantum Gravity}, vol.~19, p.~3217, 2002.

\bibitem{Astrid}
A.~Eichhorn, ``More on shape dynamics.'' ILQGS seminar.
  http://ilqgs.blogspot.com.br/2012/10/more-on-shape-dynamics.html.

\bibitem{Asymptotic_SD}
H.~Gomes, ``{Poincar\'e invariance and asymptotic flatness in Shape
  Dynamics},'' {\em ARXIV:1212.1755. Accepted for publication in Phys. Rev.
  D.}, 2012.

\bibitem{York}
J.~W. York, ``Gravitational degrees of freedom and the initial-value problem,''
  {\em Phys. Rev. Lett.}, vol.~26, pp.~1656--1658, 1971.

\bibitem{SD:FAQ}
H.~Gomes and T.~Koslowski, ``{Frequently asked questions about Shape
  Dynamics},'' {\em ARXIV:1211.5878}, 2012.

\bibitem{Henneaux&Teitelboim}
C.~T. M.~Henneaux, {\em Quantization of Gauge Systems}.
\newblock Princeton University Press, 1994.

\bibitem{Strominger}
A.~Strominger, ``{Inflation and the dS / CFT correspondence},'' {\em JHEP},
  vol.~0111, p.~049, 2001.

\end{thebibliography}

\end{document}